\documentclass[a4paper]{jpconf}
\usepackage{graphicx}
\newcommand{\Tc}{T_{\mbox{\scriptsize C}}}
\newcommand{\TN}{T_{\mbox{\scriptsize N}}}

\begin{document}
\title{Anomalous phase of MnP at very low field}

\author{T Yamazaki$^1$, Y Tabata$^1$, T Waki$^1$, H Nakamura$^1$, M Matsuura$^2$, and N Aso$^3$}

\address{$^1$Dept. of Materials Science and Engineering, Kyoto University, Kyoto 606-8501, Japan}
\address{$^2$Dept. of Earth and Space Science, Osaka University, Osaka 560-0043, Japan}
\address{$^3$Dept. of Engineering and Science, Ryukyu University, Okinawa 960-0213, Japan}

\ead{t.yamazaki@ky4.ecs.kyoto-u.ac.jp}

\begin{abstract}
Manganese phosphide MnP has been investigated for decades because of its rich magnetic phase diagram.
It is well known that the MnP exhibits the ferromagnetic phase transition at $\Tc=292$ K and the helical magnetic phase below $\TN=47$ K at zero field.
Recently, a novel magnetic phase transition was observed at $T^* = 282$ K when the magnetic field is lower than 5 Oe.
However, the nature of the new phase has not been illuminated yet.
In order to reveal it, we performed the AC and the DC magnetization measurements for a single crystal MnP at very low field.
A divergent behavior of the real and the imaginary part of the AC susceptibility and a sharp increase of  the DC magnetization was observed at $T^*$, indicating the magnetic phase transition at $T^*$.
Furthermore a peculiar temperature hysteresis was observed: namely, the magnetization depends on whether cooling sample
 to the temperature lower than $\TN$ or not before the measurements.
This hysteresis phenomenon suggests the complicated nature of the new phase and
a strong relation between the magnetic state of the new phase and the helical structure. 
\end{abstract}

%\section{Introduction}
Manganese phosphide MnP has been investigated by a lot of researchers from 1960's  because of its richness of the magnetic phases and the Lifshits critical behavior in magnetic field \cite{becerra2,shapira1,moon1,bindilatti}.
The crystal structure of MnP is a distorted orthorhombic one and its space group is Pbnm, where the lattice parameters are $a=5.92 \mbox{\AA}, b=5.26 \mbox{\AA}$ and $c=3.17 \mbox{\AA}$ at room temperature.
The c-axis is the direction of easy magnetization. The b- and a-axis are the intermediate- and hard-magnetization directions, respectively.
In zero external field, the ferromagnetic phase transition is undergone at $\Tc = 292$ K, where the spins are parallel to the c-axis \cite{Huber,felcher}. 
Below $\TN=47$ K, it transforms into the helical structure with the magnetic propagation vector $q_{mag} = (0.117,0,0)$, in which the spins lie in bc-plane \cite{felcher,forsyth,obara}.

The phase transitions of MnP can be simply explained in terms of the competition between ferromagnetic and antiferromagnetic interactions \cite{dobrzynski,yoshizawa} and it has been considered that its magnetic properties were almost understood by the extensive investigations for a long period. 
The novel transition was , however, observed at $T^*=282$ K by Becerra \cite{becerra1}.
He measured the AC susceptibility with the AC- and the DC-field along the b-axis 
and found a sharp increases of the susceptibility at $T^{\ast}$ when the DC-field is lower than 5 Oe. 
He suggested that the transition at $T^*$ is a spin reorient transition; namely,  the spins slightly incline toward the b-direction from the c-direction below $T^{\ast}$. 
The angle of inclination become lower with decreasing temperature and the spins go back to the c-direction around $\TN$.
However, no other experiments have been performed to illuminate the nature of the new phase.

In order to reveal it, we performed the AC and the DC magnetization measurements in very low field. 
A divergent behavior was observed in both the real and the imaginary parts of the AC susceptibility at $T^*$ with the very low AC- field of 0.5 Oe in zero static field.
Additionally, a remarkable increase of the DC magnetization at $T^*$ was observed as well. Those behavior indicate the magnetic phase transition at $T^*$.
Furthermore a peculiar temperature hysteresis was observed in both the AC and the DC measurements: namely, the AC susceptibility and the DC magnetization depend on whether cooling the sample
 to the temperature lower than $\TN$ or not before the measurements.
This hysteresis phenomenon suggests the complicated nature of the new phase and a strong relation between the magnetic state of the new phase and the helical structure. 

%\section{Experimental results}
A single crystalline sample of MnP was grown by the temperature gradient furnace technique \cite{Huber}. 
The size of the sample is $0.7\times 2.5 \times 1.0$ mm$^3$.
The AC and the DC magnetizations were measured by using a SQUID magnetometer (Quantum Design MPMS) with the applied AC- and DC-field along the b-axis.
The AC measurements were performed with the AC-field of 0.5 Oe and the frequency of 10 Hz in zero static field. 
The applied field in the DC measurements was 3 Oe.

\begin{figure}[h]
\begin{center}
	 \includegraphics[width=80mm]{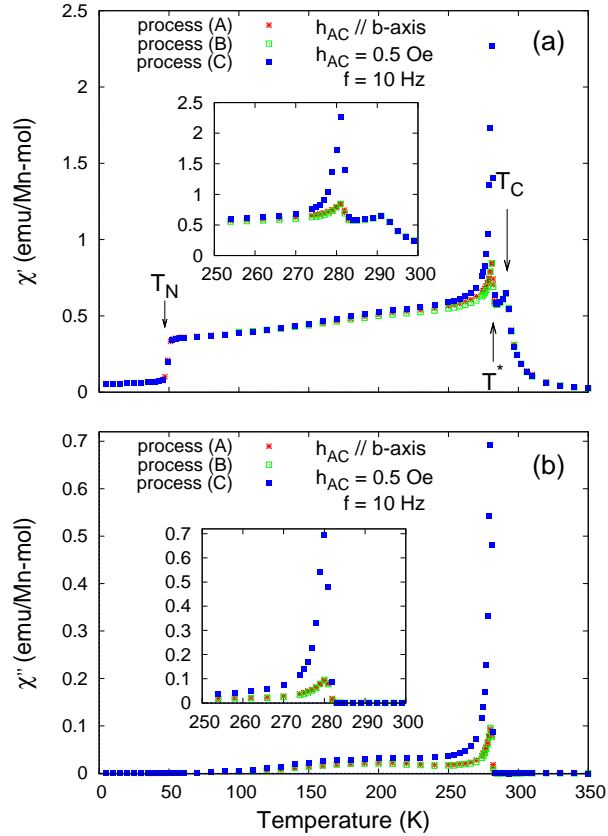}
	\caption{Temperature dependence of the AC susceptibility of MnP in the various processes. The details of the measurement processes (A)-(C) are described in the text. (a) is the real part and (b) is the imaginary part of the AC susceptibility, respectively. }
\end{center}
\end{figure}

Figure 1 shows the temperature dependences of the real and the imaginary parts of the AC susceptibility, $\chi'$ and $\chi''$, respectively.
The AC susceptibility was measured in the following three processes: (A) the cool-down process from 350 K; (B) the warm-up process after cooling the sample to 100 K which is the temperature above $\TN$; (C) the warm-up process after cooling the sample to 5 K which is the temperature below $\TN$.
The manners of the temperature dependence in the three processes are quite different in the vicinity of $T^*$.
In all processes, sharp increases of the real and the imaginary parts of the AC susceptibility are found around $T^{\ast} = 282$ K, 
however, the peaks at $T^{\ast}$ in the process (C) is much higher than those in the process (A) and (B). 
The divergent behaviors of the real and the imaginary parts of the AC susceptibility in the process (C) indicate 
the magnetic phase transition at $T = T^{\ast}$. The peak of the AC susceptibility at $T^{\ast}$ is much sharper than that observed by Becerra \cite{becerra1}
 because the AC field of 0.5 Oe in our experiments is much smaller than that in his work. 

\begin{figure}[h]
\begin{center}
	 \includegraphics[width=100mm]{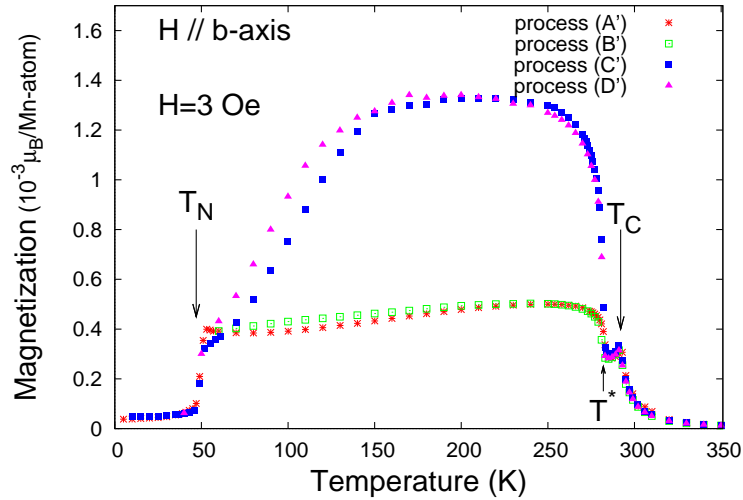}
	\caption{Temperature dependence of the DC magnetization of MnP in the various processes. The details of the measurement processes (A')-(D') are described in the text. }
\end{center}
\end{figure}

The temperature dependences of the DC magnetization are shown in Figure 2.
The measurements were performed by the following four processes
after applying the field of 3 Oe at the paramagnetic temperature 350 K: (A') the cool-down process from 350 K; (B') the warm-up process after cooling the sample to 55 K which is the temperature above $\TN$;  (C') and (D') the warm-up processes after cooling the sample to 5 K and 35 K which are the temperature below $\TN$, respectively.
A sharp increase of the magnetization at $T^*$ was observed in any processes, however the magnetization remarkably depends on a process in the temperature range of $\TN < T < T^{\ast}$, as well as the AC susceptibility does in the vicinity of $T^{\ast}$. 
The values of the magnetization in the process (C') and (D'), in which the sample has been once cooled below $\TN$, are about three times greater than those in the process (A') and (B'), in which the sample has not been cooled below $\TN$ before measurements.
On the other hand, the magnetizations in the processes (A') and (B') and those in (C') and (D') are almost same respectively. 
The magnetization-curves in the processes (A') and (B') hardly show temperature dependence 
in the whole region of the temperature range of $\TN < T < T^{\ast}$, whereas those in (C') and (D') increase 
with increasing the temperature below 140 K and are almost flat until reaching $T^{\ast}$. 
Such a process-dependence of the magnetization is only observed in the intermediate temperature range of $\TN < T < T^{\ast}$ 
and the magnetizations in all processes collapse each other in the temperature ranges of $T > T^{\ast}$ and $T < \TN$.

The observed divergence of the AC susceptibility strongly suggests that the phase transition at $T^*$ is a second order one.
The remarkable increase of the DC magnetization at $T^*$ and its saturated behavior below $T^*$ indicate an appearance of the ferromagnetic component of spin along the b-axis for $\TN<T<T^*$.

The peculiar hysteresis behavior observed in the AC susceptibility and the DC magnetization suggests 
a complicated nature of the new magnetic phase.  It is noted that the hysteresis behavior of the AC susceptibility 
was observed in zero external static field. Hence, the hysteresis cannot be explained in terms of the alignment 
of the magnetic domain of the ferromagnetic b-component by the magnetic field. The DC magnetization and the AC susceptibility 
depend on only whether the sample is cooled below $\TN$ once or not, in other words, whether the sample undergoes the helical phase once or not. 
It indicates that the stability of the ferromagnetic b-component of the ordered moment is strongly related with the low temperature helical phase 
and the magnetic state of the new phase cannot be interpreted in terms of the simple inclination of the ferromagnetic ordered moment toward the b-direction. 

The magnetic phase transitions of MnP in zero and finite field along the various directions have been explained by the competition 
of the ferromagnetic and antiferromagnetic interactions without a consideration of the Dzyaloshinsky-Moriya (DM) interaction. 
The new magnetic phase is not a simple ferromagnetic state of the b-component of the spin, as discussed above. 
Another symmetry may be broken in the new phase, which could be caused by the DM interaction. 

In conclusion, we observed the divergent behavior of the AC susceptibility and the remarkable increase of the DC magnetization at $T^*=282$ K in MnP.
Furthermore the peculiar temperature hysteresis was observed  in both the AC and the DC measurements.
The hysteresis indicates a strong relation between the magnetic state of the new phase and the lower temperature helical structure 
and the possibility of another symmetry breaking in MnP. 

\section*{Acknowledgments}
This research was supported by Grant-in-Aid for the Global COE Program, "International Center for Integrated Research and Advanced Education in Materials Science", from the Ministry of Education, Culture, Sports, Science and Technology of Japan.

%%%%%%%%%%%%%%%%%%%%%%%%%%%%%%%%%%%%%%%%%%%
\end{document}